\documentclass[12pt]{article}
\begin{document}
\title{ZPF, Zitterbewegung and Inertial Mass}
\author{B.G. Sidharth\\
International Institute for Applicable Mathematics \& Information Sciences\\
Hyderabad (India) \& Udine (Italy)\\
Dipartimento di Matematica e Fisica, Universita degli studi di
Udine, Via delle scienza 206, Udine 33100 (Italy)}
\date{}
\maketitle
\begin{abstract}
In this paper, we argue that the background Zero Point Field,
induces an oscillatory motion which we identify with zitterbewegung.
We deduce that this is the origin of inertial mass.
\vspace{5 mm}
\begin{flushleft}
\end{flushleft}
\end{abstract}
\section{Introduction}
We would first like to point out that a background Zero Point Field
of the kind we have been considering can explain the Quantum
Mechanical spin half (as also the anomalous $g = 2$ factor) for an
otherwise purely classical electron \cite{sachi,uof,milonniqv}. The
key point here is (Cf.ref.\cite{sachi}) that the classical angular
momentum $\vec r \times m \vec v$ does not satisfy the Quantum
Mechanical commutation rule for the angular momentum $\vec J$.
However when we introduce the background Zero Point Field (ZPF), the
momentum now becomes
\begin{equation}
\vec J = \vec r \times m \vec v + (e/2c) \vec r \times (\vec B
\times \vec r) + (e/c) \vec r \times \vec A^0 ,\label{3ez5}
\end{equation}
where $\vec A^0$ is the vector potential associated with the ZPF and
$\vec B$ is an external magnetic field introduced merely for
convenience, and which can be made vanishingly
small.\\
It can be shown that $\vec J$ in (\ref{3ez5}) satisfies the Quantum
Mechanical commutation relation for $\vec J \times \vec J$. At the
same time we can deduce from (\ref{3ez5})
\begin{equation}
\langle J_z \rangle = - \frac{1}{2} \hbar
\omega_0/|\omega_0|\label{3ez6}
\end{equation}
Relation (\ref{3ez6}) gives the correct Quantum Mechanical results referred to above.\\
From (\ref{3ez5}) we can also deduce that
\begin{equation}
l = \langle r^2 \rangle^{\frac{1}{2}} =
\left(\frac{\hbar}{mc}\right)\label{3ez7}
\end{equation}
Equation (\ref{3ez7}) shows that the mean dimension of the region in
which the ZPF fluctuation contributes is of the order of the Compton
wavelength of the electron. By relativistic covariance
(Cf.ref.\cite{uof}), the corresponding time scale is at the Compton
scale. Dirac, in his relativistic theory of the electron encountered
the zitterbewegung effects within the Compton scale \cite{diracpqm}
and he had to invoke averages over this scale to recover meaningful
results. We on the other hand have shown elsewhere, without invoking
the ZPF \cite{bgsfpl1552002} how spin follows, that is we get
(\ref{3ez6}) and (\ref{3ez7}) using zitterbewegung conclusions. We
would next like to show that the ZPF leads to the mass, that is,
inertial mass of a particle.
\section{Modelling the ZPF}
We can model the ZPF in terms of a Weiner process \cite{tduniv}. In
this case the right hand and left hand time derivatives at any
instant are unequal. Let us push these considerations further. We
would like to point out that it would be reasonable to expect that
the Weiner process is related to the ZPF which is the Zero Point
Energy of a Quantum Harmonic oscillator. We can justify this
expectation as follows: Let us denote the forward and backward time
derivatives with respect to time by $d_+$ and $d_-$. In usual theory
where time is differentiable, these two are equal, but we have on
the contrary taken them to be unequal. Let us consider the simple
case,
\begin{equation}
d_- = a - d_+\label{2ae}
\end{equation}
Then we have from Newton's second law in the absence of forces,
\begin{equation}
\ddot{x} + k^2 x = a\dot{x}\label{2be}
\end{equation}
wherein the new nondifferentiable effect (\ref{2ae}) is brought up.
In a normal vacuum with usual derivatives and no external forces,
Newtonian Mechanics would give us instead the equation
\begin{equation}
\ddot{x} = 0\label{2ce}
\end{equation}
A comparison of (\ref{2be}) and (\ref{2ce}) shows that the Weiner
process converts a uniformly moving particle, or a particle at rest
into an oscillator-- a damped oscillator, strictly speaking. Indeed
in (\ref{2be}) if we take as a first approximation
\begin{equation}
\dot{x} \approx \langle \dot{x} \rangle = 0\label{2de}
\end{equation}
then we would get the exact oscillator equation
\begin{equation}
\ddot{x} + k^2 x = 0\label{2ee}
\end{equation}
for which in any case, consistently (\ref{2de}) is correct.
Moreover, the ZPF results from the quantized oscillator
(\ref{2ee}). We will return to this a little later.\\
We can push these considerations even further and
deduce alternatively, the Schrodinger equation, as pointed out by
Nottale \cite{nottalefractal}. The genesis of Special Relativity too
can be found in the Weiner process (Cf.ref.\cite{bgsfpl1552002} for
a detailed discussion). Let us examine this
more closely.\\
We first define a complete set of base states by the subscript
$\imath \quad \mbox{and}\quad U (t_2,t_1)$ the time elapse operator
that denotes the passage of time between instants $t_1$ and $t_2$,
$t_2$ greater than $t_1$. We denote by, $C_\imath (t) \equiv <
\imath |\psi (t)
>$, the amplitude for the state $|\psi (t) >$ to be in the state $|
\imath >$ at time $t,$ and \cite{ijpap,cu}
$$< \imath |U|j > \equiv U_{\imath j}, U_{\imath j}(t + \Delta t,t) \equiv
\delta_{\imath j} - \frac{\imath}{\hbar} H_{\imath j}(t)\Delta t.$$
We can now deduce from the super position of states principle that,
\begin{equation}
C_\imath (t + \Delta t) = \sum_{j} [\delta_{\imath j} -
\frac{\imath} {\hbar} H_{\imath j}(t) \Delta t] C_j (t)\label{2xe}
\end{equation}
and finally, in the limit,
\begin{equation}
\imath \hbar \frac{dC_\imath (t)}{dt} = \sum_{j} H_{\imath
j}(t)C_j(t)\label{2fe}
\end{equation}
where the matrix $H_{\imath j}(t)$ is identified with the
Hamiltonian operator. It must be mentioned that $H_{\imath j}$ gives
the probability for a state $C_\imath$ to be found in state $C_j$.
Such a non-local transition is allowed within the Compton scale, as
discussed in detail by Weinberg \cite{weinberggrqc}. As Weinberg
notes:''Although the relativity of temporal order raises no problems
for classical physics, it plays a profound role in quantum theories.
The uncertainty principle tells us that when we specify that a
particle is at position $x_1$ at time $t_1$, we cannot also define
its velocity precisely. In consequence there is a certain chance of
a particle getting from $x_1$ to $x_2$ even if $x_1 - x_2$ is
space-like, that is, $| x_1 - x_2 | > |x_1^0 - x_2^0|$. To be more
precise, the probability of a particle reaching $x_2$ if it starts
at $x_1$ is nonnegligible as long as
$$ 0 \leq (x_1 - x_2)^2 -
(x_1^0 - x_2^0)^2 \leq \frac{\hbar^2}{m^2} \cdots "$$ We have argued
earlier at length that (\ref{2fe}) leads to the Schrodinger equation
\cite{ijpap,cu}. In this derivation, the mass term in the
Schrodinger equation comes from $H_{\imath j}$: The matrix $H(x,x')$
gives the probability amplitude for the particle at $x$ to be found
at $x'$, that is,
\begin{equation}
H(x,x') = < \psi (x')|\psi (x) >\label{e13}
\end{equation}
where as is usual we write $C(x)\equiv \psi (x)(\equiv | \psi (x)
>)$,
the state of a particle at the point $x$).\\
Usually the amplitude $H(x,x')$ is non-zero only for neighbouring
points $x$ and $x'$, that is, $H(x,x') = f(x)\delta (x-x').$ But if
$H(x,x')$ is not of this form, then there is a non-zero amplitude
for the particle to "jump" to an other than neighbouring point. In
this case $H(x,x')$ may be described as a non local amplitude. However, in the light
of the above, we consider such a non local behaviour, only within the Compton
scale.\\
The contribution of this term is,
$$\int \psi^* (x')\psi (x) \psi (x') U(x')dx'$$
where,\\
i)$U(x) = 1$ for $|x| < R,R$ arbitrarily large and also $U(x)$ falls
off rapidly as $|x| \to \infty; U(x)$ has been introduced merely to
ensure
the convergence of the integral; and\\
ii)$H(x,x') = < \psi (x')\psi (x) > = \psi^* (x')\psi (x).$\\
The presence of the, what at first sight may seem troublesome,
non-linear and non-local term above.\\
Finally we get
\begin{equation}
m_0 = \int \psi^* (x') \psi (x')U(x')dx'\label{e17}
\end{equation}
(Cf.\cite{cu} for details).\\
In the above we have taken the usual unidirectional time to deduce
the non relativistic Schrodinger equation. If however we consider a
Weiner process in (\ref{2xe}) then we will have to consider instead
of (\ref{2fe})
\begin{equation}
C_\imath (t - \Delta t) - C_\imath (t + \Delta t) = \sum_{j}
\left[\delta_{\imath j} - \frac{\imath}{\hbar} H_{\imath j}(t)
\Delta t\right] C_j(t)\label{2ge}
\end{equation}
Equation (\ref{2ge}) in the limit can be seen to lead to the
relativistic Klein-Gordon equation rather than the Schrodinger
equation with the mass term \cite{bgscsfqfst}. Furthermore, the
Klein-Gordon equation describes the normal mode vibrations of
Harmonic Oscillators--that is, we recover (\ref{2ee}), again. Thus
the ZPF modelled by a Weiner process leads to the mass term at the
Compton scale, and indeed, special relativity. In other words, the
mass arises due to the "bootstrapping" effect within the Compton
scale that arises due to the ZPF or zitterbewegung.
\section{An Alternative Derivation}
To look at the above from a different point of view, let us start
with the Langevin equation in the absence of external
forces,\cite{reif,balescu}
$$m \frac{dv}{dt} = -\alpha v + F'(t)$$
where the coefficient of the frictional force is given by Stokes's
Law \cite{joos}
$$\alpha = 6\pi \eta a$$
$\eta$ being the coefficient of viscosity, and where we are
considering a sphere of radius $a$.\\
For $t$, we consider in the above spirit of equation (\ref{3ez7}),
that there is a cut off time $\tau$. It is known (Cf.\cite{reif})
that there is a characteristic time constant of the system, given by
$$\frac{m}{\alpha} \sim \frac{m}{\eta a},$$
so that, from Stokes's Law, as
$$\eta = \frac{mc}{a^2} \, \mbox{or}\, m = \eta \frac{a^2}{c}$$
we get
$$\tau \sim \frac{ma^2}{mca} = \frac{a}{c},$$
that is $\tau$ is the Compton time.\\
The expression for $\eta$ which follows from the fact that
$$F_x = \eta (\Delta s) \frac{dv}{dz} = m\dot v = \eta \frac{a^2}{c} \dot v ,$$
shows that the intertial mass is due to a type of "viscosity" of the
background Zero Point Field (ZPF). (Cf. also ref.\cite{rueda}).\\
To push these small scale considerations further, we have, using the
Beckenstein radiation equation\cite{ruffinizang},
$$t \equiv \tau = \frac{G^2m^3}{\hbar c^4} = \frac{m}{\eta a} = \frac{a}{c}$$
which gives
$$a = \frac{\hbar}{mc} \quad \mbox{if} \quad \frac{Gm}{c^2} = a$$
In other words the Compton wavelength equals the Schwarzchild
radius, which automatically gives us the Planck mass. Thus the
inertial mass is thrown up in these considerations at the Planck scale.\\
In fact if we now use the Langevin equation in a viscous medium
\cite{bhtd,balescu} then as the viscosity becomes vanishingly small,
it turns out that the Brownian particle moves according to Newton's
first law, that is with a constant velocity. Moreover this constant
velocity is given by (Cf.refs. \cite{bhtd,balescu}), for any mass
$m$,
\begin{equation}
\langle v^2 \rangle = \frac{kT}{m}\label{4e1a}
\end{equation}
We can derive that this velocity is that of light in a heuristic
way. We have already seen that the ZPF causes a harmonic motion. Let
us assume that the particle has a small charge $e$, just to couple
it to the ZPF. The equation of motion is now given by
(Cf.\cite{sachira})
$$\ddot{x} + \omega^2 J = (e/m) E^0_x$$
along the $x$ axis, where, suppressing the polarization states for
the moment, the random field $\vec{E}$ is given by
$$\vec{E} \propto \int d^3 \vec{K} exp [-\imath (\omega t-K\cdot \vec{r})]
\vec{a}_K + c\cdot c\cdot$$ where, owing to the randomness in phase,
their averages vanish. What this means is that, finally,
\begin{equation}
L^2 = \langle x^2 \rangle = \hbar /(m \omega ) \, \mbox{and}\,
\langle \dot{x^2}\rangle = \left(\hbar /m\right) \omega =
v^2\label{4xa}
\end{equation}
In (\ref{4xa}) above, the frequency is given by,
$$\omega = mc^2 /\hbar$$
Whence, within the Compton length,
$$\langle \dot{x}^2 \rangle = c^2$$
Using this result in (\ref{4e1a}), we get for the Planck mass, the
correct radiation for the Beckenstein temperature. For an elementary
particle in general, $T$ would be the well known Hagedorn
temperature
\cite{sivaramam}.\\
We finally make the following remark. We have seen that a particle
in the ZPF modeled by a Random electromagnetic field, leads to
equation (\ref{4xa}), viz.,
$$L^2 = \langle x^2 \rangle = \hbar / m\omega ,$$
$$\langle \dot{x^2}\rangle = (\hbar / m) \omega ,$$
where the frequency is given $\omega = mc^2 / \hbar$. In the
equation leading to (\ref{4xa}), we could also include a term which
gives the third derivative of $x$, this corresponding to the Schott
term of classical electrodynamics \cite{rohr}. However the
contribution of this term, which was introduced for energy
conservation to compensate the radiation loss of an accelerated
charge \cite{rohr} in the classical electron theory, is of the order
of the Compton wavelength \cite{iaad} and does not effect the
conclusion.\\
The above show that there is oscillation of the particle within the
Compton wavelength $L$, with velocity $c$. This as mentioned, models
the well known zitterbewegung of Dirac where also, the electron has
the speed of light within the Compton wavelength. However, what all
this means is that via the Compton length $L$, we get the inertial
mass of the particle, which is now seen to be due to the energy of
this oscillation.\\
{\bf A final comment.} Following Wheeler, we can consider the ZPF in
terms of the fluctuations of the vacuum electromagnetic field. If
this field vanishes everywhere except in a region of dimension $l$,
then the energy is given by
$$\mbox{Energy} \, \sim \hbar c/l$$
If $l$ is the Compton wavelength, this turns out to be $mc^2$
\cite{tduniv}. This reconfirms our above conclusion.

\end{document}